\documentclass[prd,twocolumn]{revtex4}

\pdfoutput=1

\usepackage{graphicx}
\usepackage{amssymb,amsmath,bm}  

\usepackage{lipsum}

\usepackage[usenames,dvipsnames]{xcolor}
\usepackage{hyperref}
\hypersetup{
    colorlinks = true,
    citecolor = {MidnightBlue},
    linkcolor = {BrickRed},
    urlcolor = {BrickRed}
}
\usepackage{lipsum}  
\usepackage{hyperref}

\newcommand{\ba}{\begin{eqnarray}}
\newcommand{\ea}{\end{eqnarray}}

\begin{document}
\title{Fast computation of angular power spectra and covariances of high-resolution cosmic microwave background maps using the Toeplitz approximation}
\author{ Thibaut Louis$^1$, Sigurd~Naess$^{2}$, Xavier Garrido$^1$, Anthony Challinor$^{3,4,5}$
             }
            \affiliation{$^{1}$Universit\'e Paris-Saclay, CNRS/IN2P3, IJCLab, 91405 Orsay, France\\
             		   $^{2}$Center for Computational Astrophysics, Flatiron Institute, 162 5th Avenue, New York, NY, USA 10010\\
  $^{3}$Institute of Astronomy, Madingley Road, Cambridge CB3 0HA, UK\\
  $^{4}$Kavli Institute for Cosmology Cambridge, Madingley Road, Cambridge CB3 0HA, UK\\
 $^{5}$DAMTP, Centre for Mathematical Sciences, Wilberforce Road, Cambridge CB3 0WA, UK}

\begin{abstract}
We present a simple approximation that can speed up the computation of the mode-coupling matrices, which are usually the bottleneck for computing unbiased angular power spectra, as well as their associated covariance matrices, of the cosmic microwave background temperature and polarization anisotropies. The approximation results in the speed up of the MASTER algorithm by more than an order of magnitude with very little
loss of precision. We demonstrate the performance on simulations of forthcoming
cosmic microwave background surveys such as the Simons Observatory and CMB-S4 for a wide variety of survey window functions. 
\end{abstract}

  \date{\today}
  \maketitle

\section{Introduction}\label{sec:intro}

Ground-based cosmic microwave background (CMB) surveys such as the Atacama Cosmology Telescope \cite{2014JCAP...10..007N, 2017JCAP...06..031L} and the South Pole Telescope \cite{2019arXiv191005748S} are producing arcminute-scale maps of the microwave sky. In the near future the large-aperture telescope of the Simons Observatory \cite{2019JCAP...02..056A, 2018SPIE10708E..04G} is expected to observe half the sky in six frequency bands, four of them with arcminute resolution, for five years.

Extracting cosmological information from these maps requires computing multiple angular power spectra and their associated covariance matrices up to multipoles $\ell_{\text{max}} \approx 10^{4}$ and is computationally challenging. The power spectra are also often used to check  the data consistency against potential systematics errors, for example by forming different null tests between subsets of the data, resulting in a large amount of extra computation.  

While many different algorithms can be used to estimate unbiased power spectra,  the standard algorithm for the estimation of power spectra  of high-resolution maps is the MASTER algorithm \cite{2002ApJ...567....2H}, which can be decomposed into four main steps.
First the spherical harmonic transforms (SHTs) of the products of the maps and some heuristic weighting functions need to be computed. These weighting functions are often the survey mask, with some suitable apodization, but may also include some additional weighting to account for anisotropic noise, for example. We shall refer to these weighting functions as the \emph{survey window functions}. Second, the \emph{pseudo}-power spectra are computed by appropriate cross-correlations of the transforms of the weighted maps. The signal part of these spectra is biased due to the coupling of the different spherical harmonics through the survey window function. This multiplicative bias is accounted for in the third step by calculating the power spectrum coupling matrices. Finally the covariances of the different power spectra are assessed. This work focuses on reducing the computational complexity of the last two steps, which scale as $\mathcal{O}(\ell^{3}_{\rm max})$ in the standard MASTER implementation.

This paper is structured as follows. In Sec.~\ref{sec:approx} we describe the MASTER algorithm and propose approximations to speed up significantly the computation of the mode-coupling matrices. In
Sec.~\ref{sec:results}, we discuss the speed of the algorithm and its precision with respect to the noise level of forthcoming CMB experiments such as the Simons Observatory \cite{2019JCAP...02..056A} and CMB-S4 \cite{2016arXiv161002743A}.
 We conclude in Sec.~\ref{sec:conclu}. 
 
All software used in this analysis is publicly available. The power spectrum code {\it pspy} can be obtained at \footnote{https://github.com/simonsobs/pspy} and the scripts used for testing the approximations at \footnote{https://github.com/thibautlouis/coupling\_approx}.

\section{Approximating the MASTER algorithm}\label{sec:approx}

In this section, we start with a short summary of the standard MASTER algorithm, introducing the different coupling matrices that need to be computed in order to recover unbiased power spectra. We also discuss how the coupling kernels enter into analytic approximations for the power spectrum covariances. We then describe the approximations that we are proposing and illustrate their behavior for a baseline window function.

\subsection{Standard  MASTER algorithm }

A simplistic data model for the temperature observed on the sky at frequency $\nu$ can be written 
\ba
\tilde{T}^{\nu}(\hat{\boldsymbol{n}})=  w^{\nu}(\hat{\boldsymbol{n}}) [b^{\nu}(\hat{\boldsymbol{n}}) * T^{\nu}(\hat{\boldsymbol{n}}) + n^{\nu}(\hat{\boldsymbol{n}}) ],
\ea
where  $w^{\nu}(\hat{\boldsymbol{n}})$ is the survey window function, $b^{\nu}(\hat{\boldsymbol{n}})$ is the (circularly-symmetric) instrument beam convolving the signal from the sky, $T^\nu(\hat{\boldsymbol{n}})$, and $ n^{\nu}(\hat{\boldsymbol{n}})$ represents the noise on the observation. The multiplication by a window function leads, in harmonic space, to a coupling between different multipoles
\ba  
\tilde{a}^{T,\nu}_{\ell m} = \sum_{\ell' m'} (b^{\nu}_{\ell'} a^{T,\nu}_{\ell' m'}+  n^{\nu}_{\ell' m'}) \int {\rm d}^2 \hat{\boldsymbol{n}} Y_{\ell' m'}(\hat{\boldsymbol{n}}) w^{\nu}(\hat{\boldsymbol{n}}) Y^{*}_{\ell m}(\hat{\boldsymbol{n}}) , \nonumber \\
\ea 
where $a^{T,\nu}_{\ell m}$ are the true spherical multipoles of the temperature field $T^\nu(\hat{\boldsymbol{n}})$, and we have also decomposed the noise into spherical multipoles $n^\nu_{\ell m}$. The spherical multipoles of the instrument beam are
$\sqrt{(2\ell+1)/4\pi} b^\nu _{\ell} \delta_{m0}$.
The effect of this coupling can be written as a linear operation on the power spectrum. 
We consider the typical case of a cross-power spectrum between maps with independent noise (e.g., maps obtained at different epochs or at different frequencies).
For such maps at frequencies $\nu_{1}$ and $\nu_{2}$, the pseudo-power spectrum,
\ba
\tilde{C}_\ell^{\nu_1 \nu_2} \equiv \frac{1}{2\ell+1} \sum_m \tilde{a}^{T,\nu_1}_{\ell m} 
\left(\tilde{a}^{T,\nu_2}_{\ell m}\right)^* , 
\ea
has expectation value 
\ba
\langle \tilde{C}^{\nu_{1}\nu_{2}}_{\ell_{1}} \rangle  = \sum_{\ell_2}M^{ \nu_{1}\nu_{2}}_{ \ell_1 \ell_2}  C^{\nu_{1}\nu_{2}}_{\ell_2},
\ea
where $C^{\nu_{1}\nu_{2}}_{\ell_2}$ is the true underlying signal power spectrum (we assume that the $T^\nu(\hat{\boldsymbol{n}})$ are statistically isotropic), $M^{ \nu_{1}\nu_{2}}_{ \ell_{1} \ell_{2}}$ is the mode-coupling matrix, which depends only on the instrument beams and the power spectrum of the survey window functions.
Defining $F^{ \nu_{1}\nu_{2}}_{\ell_{2}} = (2\ell_2 +1)b^{\nu_{1}}_{\ell_{2}}b^{\nu_{2}}_{\ell_{2}} $,  the mode-coupling matrix can be written 
\ba
\label{eq:coupling}
M^{ \nu_{1}\nu_{2}}_{ \ell_1 \ell_2} &=&  F^{ \nu_{1}\nu_{2}}_{\ell_{2}}   \sum_{ \ell_3} \frac{(2\ell_3+1)}{4\pi} {\cal W}^{\nu_{1}\nu_{2}}_{\ell_3} 
\left(\begin{array}{clcr}
\ell_1 & \ell_2 & \ell_3\\
0 & 0 & 0 \end{array}\right)^{2} \nonumber \\
&=& F^{ \nu_{1}\nu_{2}}_{\ell_{2}}  \Xi^{00}_{\ell_{1}  \ell_{2}}(  w^{\nu_{1}},  w^{\nu_{2}}).
\ea 
Here ${\cal W}^{\nu_{1}\nu_{2}}_{\ell_3} = {\cal W}_{\ell_3}(  w^{\nu_{1}},  w^{\nu_{2}})$ is the cross power spectrum of the window functions of the maps at frequencies $\nu_{1}$ and $\nu_{2}$.
We can use $M^{ \nu_{1}\nu_{2}}_{ \ell_1 \ell_2} $ to form an unbiased estimator of the true power spectrum  
\ba
\hat{C}^{\nu_{1}\nu_{2}}_{\ell_{1}}   = \sum_{\ell_2}(M^{-1})^{ \nu_{1}\nu_{2}}_{ \ell_1 \ell_2}  \tilde{C}^{\nu_{1}\nu_{2}}_{\ell_2} .
\ea 
This is the basis of the standard MASTER algorithm. 

Analytic covariances matrices of the power spectra can also be computed \cite{2006MNRAS.370..343E, 2005MNRAS.360.1262B}
\ba
{\rm Cov} ( {\bm \hat{C}}^{\nu_{1}\nu_{2}},{\bm \hat{C}}^{\nu_{3}\nu_{4}} ) =  ({\bm M}^{ \nu_{1}\nu_{2}})^{-1}  {\bm V^{ \nu_{1}\nu_{2} \nu_{3}\nu_{4}}} ( {\bm M}^{ \nu_{3}\nu_{4}})^{-1, T} .
\ea
Evaluating the matrix $ {\bm V}$ would require naively $ {\cal O}( \ell^{6}_{\rm max})$ operations, however a good approximation to this matrix has been found with the following expression 
\ba
\label{eq:cov}
V^{ \nu_{1}\nu_{2} \nu_{3}\nu_{4}}_{\ell_{1} \ell_{2}} &\approx& C^{\nu_{1} \nu_{3}}_{\ell_{1} \ell_{2}} C^{\nu_{2} \nu_{4}}_{\ell_{1} \ell_{2}}  \Xi^{00}_{\ell_{1} \ell_{2}}(  w^{\nu_{1}}w^{\nu_{3}},  w^{\nu_{2}} w^{\nu_{4 }}) \nonumber \\
&&\mbox{} +  C^{\nu_{1} \nu_{4}}_{\ell_{1} \ell_{2}}  C^{\nu_{2} \nu_{3}}_{\ell_{1} \ell_{2}}  \Xi^{00}_{\ell_{1} \ell_{2}}(  w^{\nu_{1}}w^{\nu_{4}},  w^{\nu_{2}} w^{\nu_{3}}) ,
\ea
where $ C^{\nu_{1} \nu_{3}}_{\ell \ell'}= \sqrt{ C^{\nu_{1} \nu_{3}}_{\ell } C^{\nu_{1} \nu_{3}}_{\ell'}}$ is a symmetrized version of the beam-convolved power spectra. 
The  most important step of power spectra analysis with the MASTER algorithm is the calculation of the coupling matrices, with overall scaling $ {\cal O}( \ell^{3}_{\rm max})$. They appear both in Eqs~\eqref{eq:coupling} and~\eqref{eq:cov}. An analysis involving $n_{\rm freq} $ frequency bands, requires computing $n_{\rm spec} = n_{\rm freq} (n_{\rm freq} +1)/2$ spectra and $n_{\rm cov} = n_{\rm spec}  (n_{\rm spec} +1)/2$ covariance elements, making the full computation $ {\cal O}(n^{4}_{\rm freq}  \ell^{3}_{\rm max})$.  

The computational problem gets worse when we include polarisation data. Obtaining the full T-, E- and B-mode cross-power spectra requires computing  coupling terms of the form (e.g.,  Ref.~\cite{2019MNRAS.484.4127A})
 \begin{widetext}
\ba
\Xi^{00}_{\ell_{1}  \ell_{2}}(  w_{T}^{\nu_{1}},  w_{T}^{\nu_{2}}) &=& \sum_{\ell_{3}}  \frac{(2\ell_3+1)}{4\pi} {\cal W}_{\ell_3}(  w_{T}^{\nu_{1}},  w_{T}^{\nu_{2}}) \left(\begin{array}{clcr}
\ell_1 & \ell_2 & \ell_3\\
0 & 0 & 0 \end{array}\right)^{2}, \nonumber \\
\Xi^{++}_{\ell_{1}  \ell_{2}}(  w_{P}^{\nu_{1}},  w_{P}^{\nu_{2}}) &=& \sum_{\ell_{3}}  \frac{(2\ell_3+1)}{4\pi} {\cal W}_{\ell_3}(  w_{P}^{\nu_{1}},  w_{P}^{\nu_{2}}) \left(\begin{array}{clcr}
\ell_1 & \ell_2 & \ell_3\\
2 & -2 & 0 \end{array}\right)^{2} \frac{\left(1+ (-1)^{\ell_1+ \ell_2 + \ell_3}\right)}{2} , \nonumber \\
\Xi^{--}_{\ell_{1}  \ell_{2}}(  w_{P}^{\nu_{1}},  w_{P}^{\nu_{2}}) &=&   \sum_{\ell_{3}}  \frac{(2\ell_3+1)}{4\pi} {\cal W}_{\ell_3}(  w_{P}^{\nu_{1}},  w_{P}^{\nu_{2}}) \left(\begin{array}{clcr}
\ell_1 & \ell_2 & \ell_3\\
2 & -2 & 0 \end{array}\right)^{2} \frac{\left(1- (-1)^{\ell_1+ \ell_2 + \ell_3}\right)}{2} , \nonumber \\
\Xi^{02}_{\ell_{1}  \ell_{2}}(  w_{T}^{\nu_{1}},  w_{P}^{\nu_{2}}) &=&   \sum_{\ell_{3}}  \frac{(2\ell_3+1)}{4\pi} {\cal W}_{\ell_3}(  w_{T}^{\nu_{1}},  w_{P}^{\nu_{2}}) \left(\begin{array}{clcr}
\ell_1& \ell_2 & \ell_3\\
2 & -2 & 0 \end{array}\right) 
\left(\begin{array}{clcr}
\ell_1 & \ell_2 & \ell_3\\
0 & 0 & 0 \end{array}\right) , \nonumber \\
\Xi^{20}_{\ell_{1}  \ell_{2}}(  w_{P}^{\nu_{1}},  w_{T}^{\nu_{2}}) &=&   \sum_{\ell_{3}}   \frac{(2\ell_3+1)}{4\pi} {\cal W}_{\ell_3}(  w_{P}^{\nu_{1}},  w_{T}^{\nu_{2}}) \left(\begin{array}{clcr}
\ell_1& \ell_2 & \ell_3\\
2 & -2 & 0 \end{array}\right) 
\left(\begin{array}{clcr}
\ell_1 & \ell_2 & \ell_3\\
0 & 0 & 0 \end{array}\right) ,
\ea
 \end{widetext}
 and many more different covariance matrix elements.
Here, $\Xi^{++}$ describes the coupling of the EE pseudo-power spectrum to the true EE spectrum, and similarly for BB, while $\Xi^{--}$ describes the coupling of the EE pseudo-power to the true BB and vice-versa. The couplings $\Xi^{02}$ and $\Xi^{20}$
relate the TE pseudo-power to the true TE. If we also consider measuring the TB and EB spectra, which are expected to vanish if parity invariance is respected in the mean, their pseudo-spectra would be related to any true TB and EB power by the couplings $\Xi^{02}$ (and $\Xi^{20}$) and $\Xi^{++}-\Xi^{--}$, respectively.
Note that we have allowed for the polarisation survey window function $w_P^\nu$ to differ from the temperature one.
For multi-band, high-resolution experiments, the task of computing all these couplings becomes difficult even on modern supercomputers.

\subsection{The Toeplitz approximation}

\begin{figure*}
\includegraphics[width=0.75\textwidth]{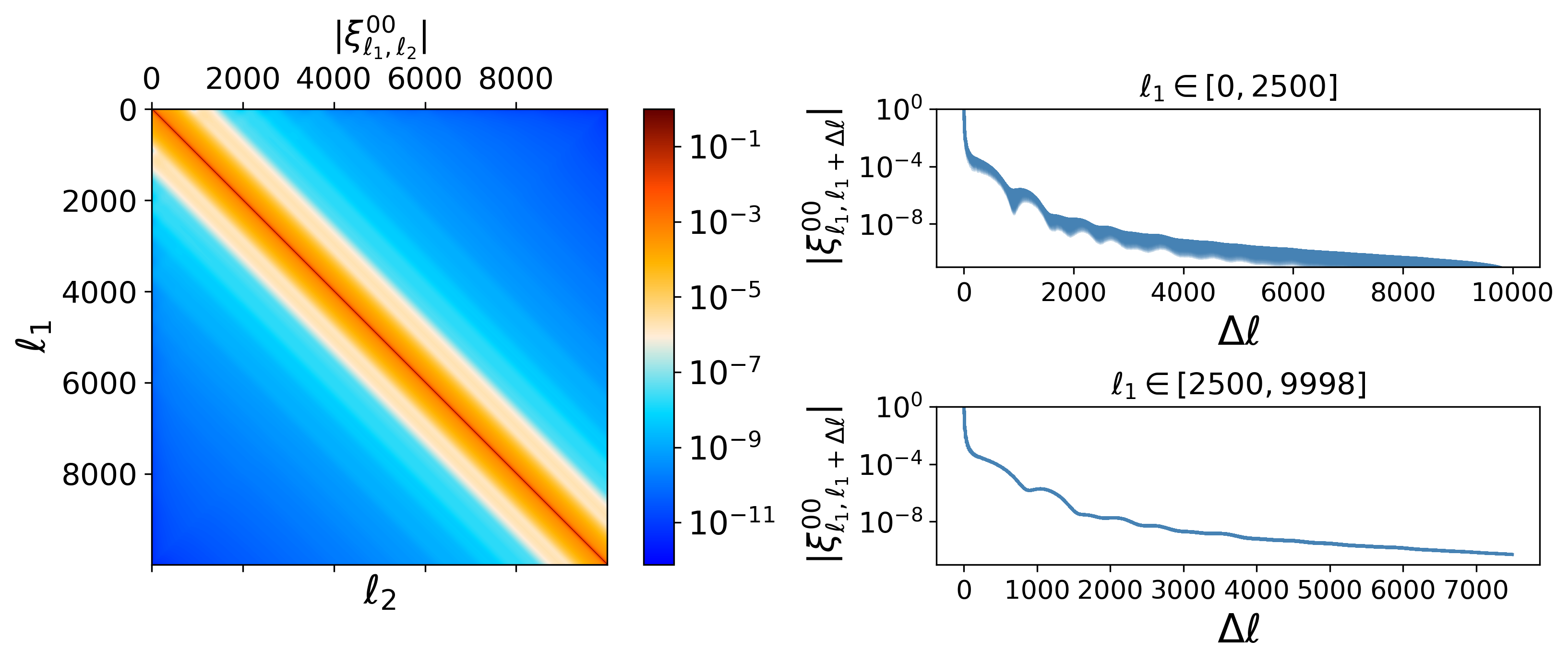}
\includegraphics[width=0.75\textwidth]{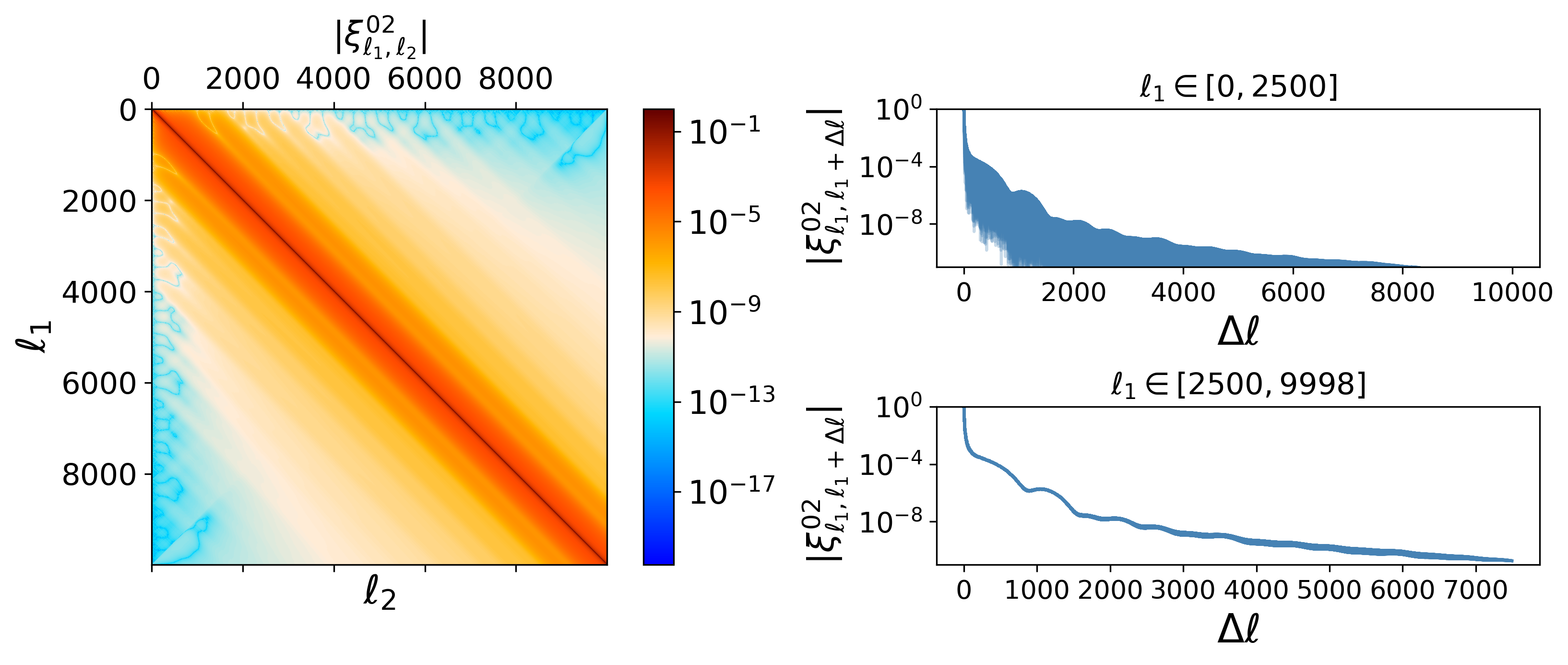}
\includegraphics[width=0.75\textwidth]{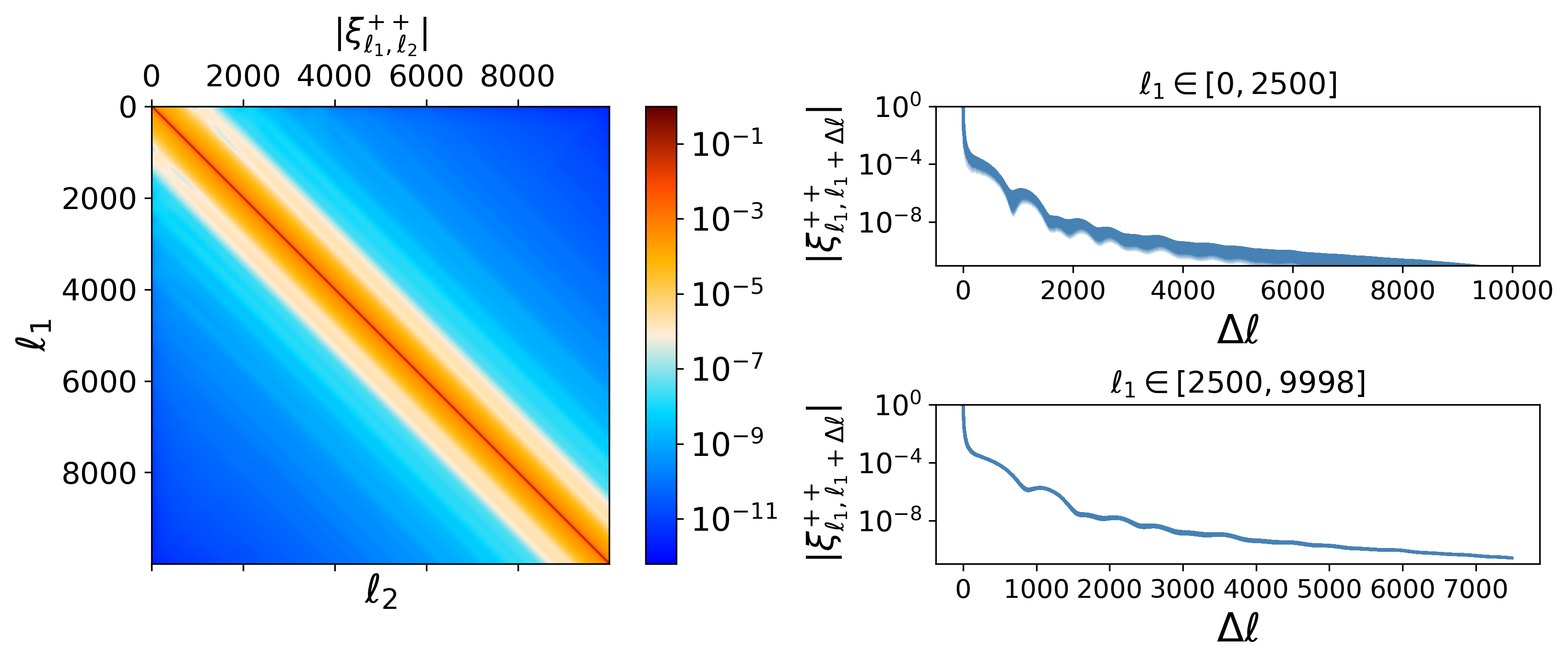}
\includegraphics[width=0.75\textwidth]{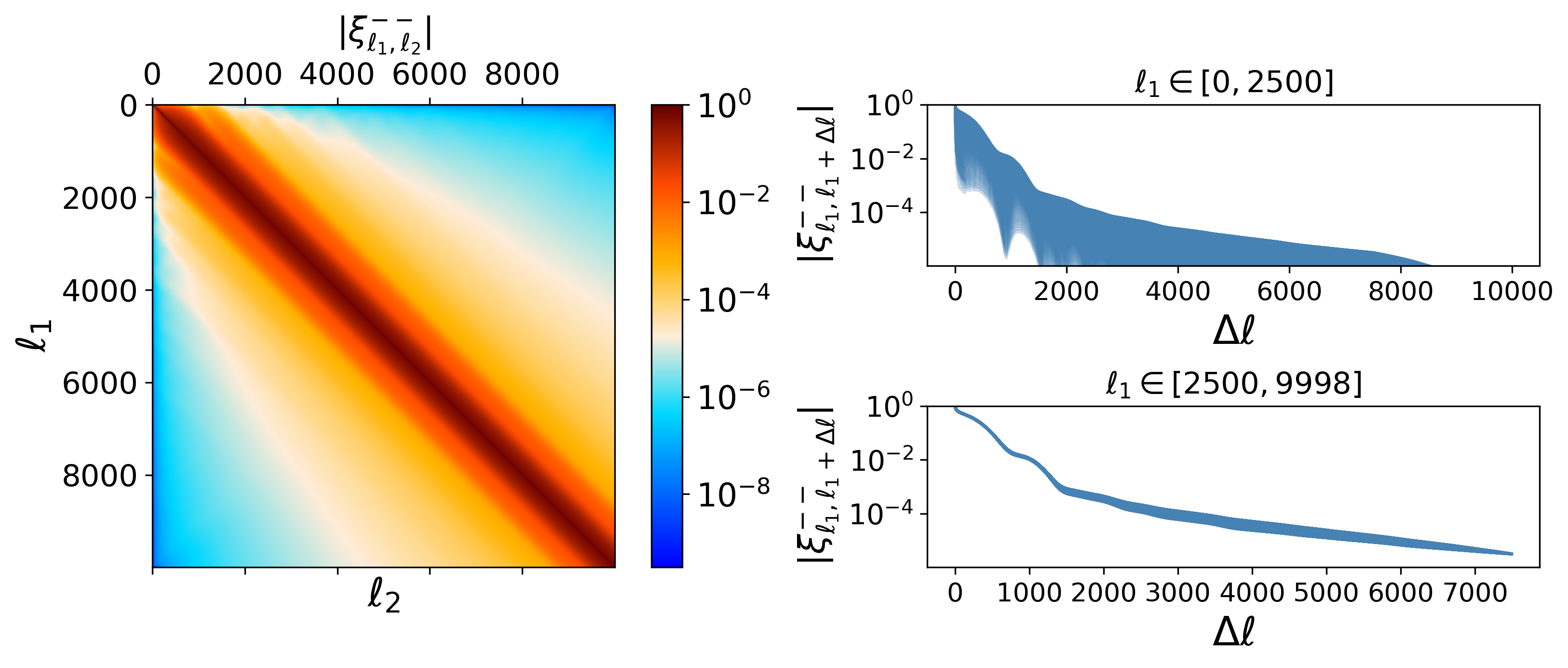}
\caption{Normalized coupling coefficients involved in computation of power spectra and covariance matrices. Left: the absolute value of the full matrices $\xi^{\alpha \beta}_{\ell_{1}  \ell_{2}}$
on a logarithmic scale. Right: 
one-dimensional plots of all rows of the matrix on the left starting from the diagonal, plotted on top of each other as blue curves. The top panels shows the curves for $\ell_{1} \leq 2500$. At these low $\ell_{1}$s the Toeplitz approximation is less accurate, and the curves spread out to form a band. The bottom panels shows the curves for $\ell_{1} \geq 2500$. In this range the Toeplitz approximation is accurate enough and the curves fall on top of each other. We note that the Toeplitz approximation is a slightly worse approximation for the $\xi^{--}_{\ell_{1}  \ell_{2}}$ matrix. However, in practice, the error on this term, representing coupling between E- and B-modes, is subdominant.}

 \label{fig:correlation}
\end{figure*}

\begin{figure*}
\includegraphics[width=0.9\columnwidth]{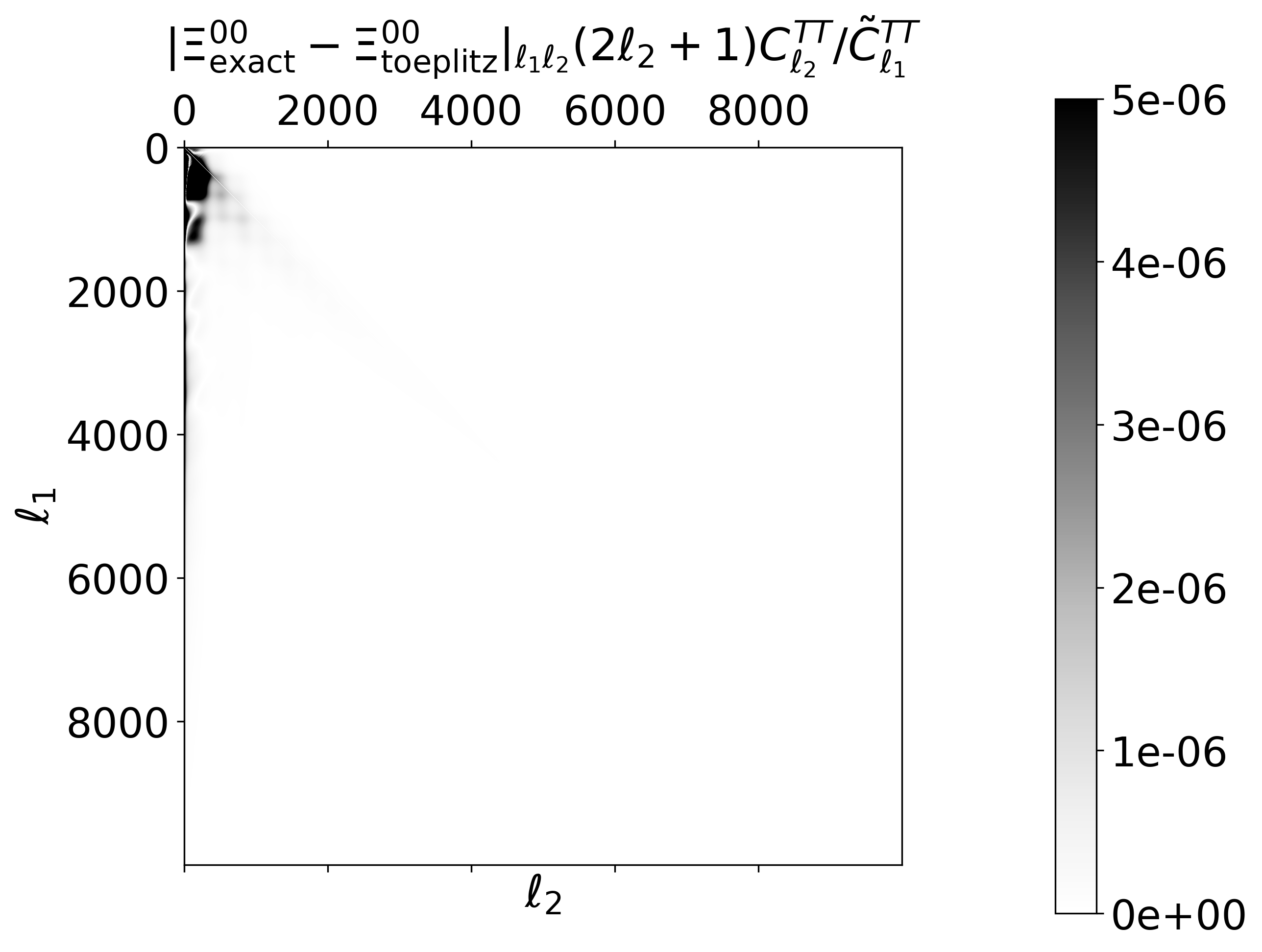}
\includegraphics[width=0.9\columnwidth]{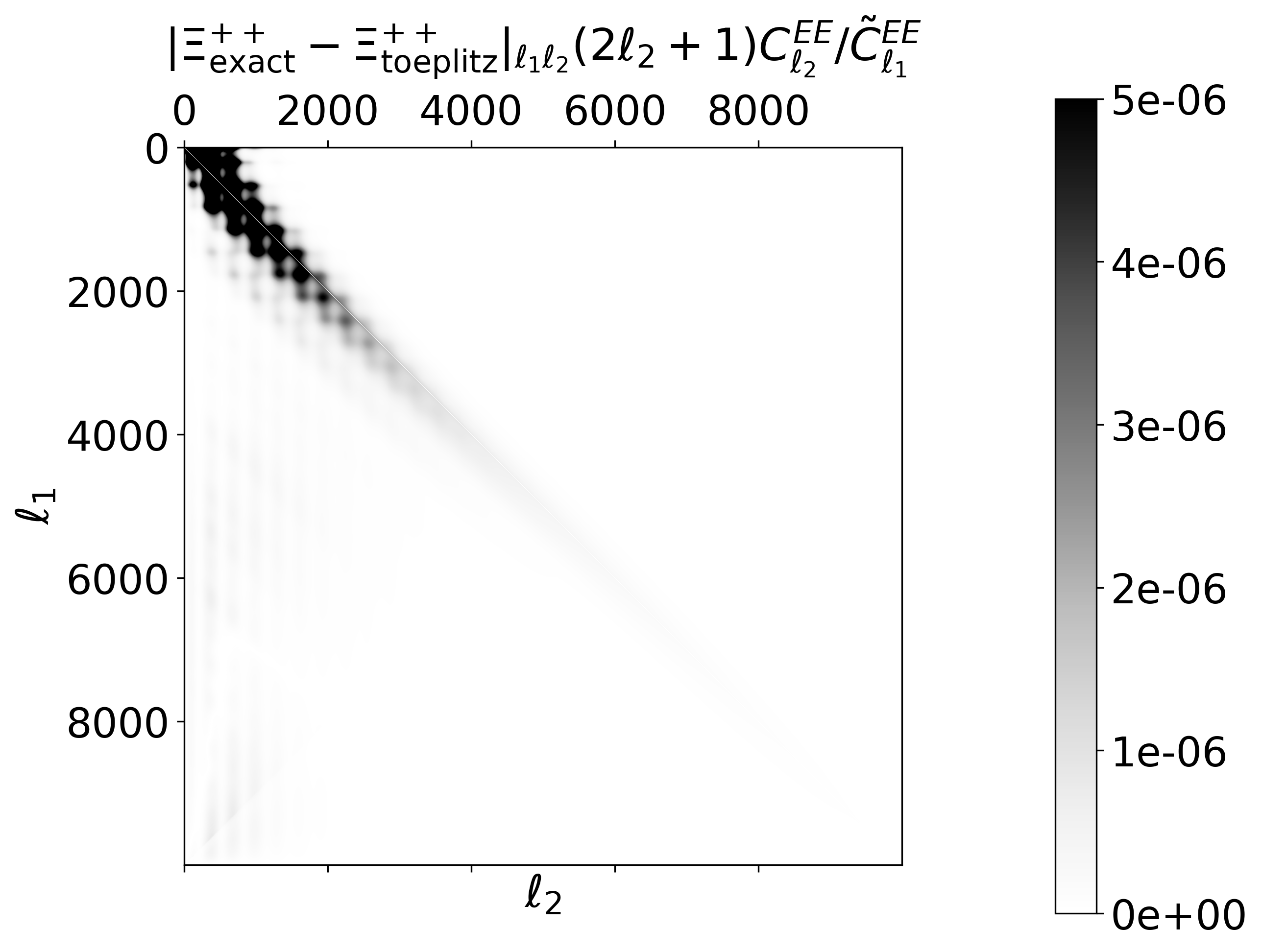}
\caption{Fractional difference  on the inferred pseudo-power spectrum between the exact computation and the Toeplitz approximation for TT (left) and EE (right). We show the magnitude of the contribution per matrix element. The approximation fails on large and intermediate scales but become very precise on small angular scales. Note that for the right panel, we neglect the effect of B-to-E leakage, assuming a vanishing contribution from  the BB power spectrum to the EE power spectrum.}
 \label{fig:metric}
\end{figure*}

It is therefore interesting to look more closely at the properties of the coupling coefficients $\Xi^{\alpha \beta}_{\ell_{1}  \ell_{2}}$   ($\alpha\beta \in \{00, 02, 20, ++, -- \})$. First, they are symmetric in $\ell_{1}, \ell_{2}$ due to the symmetry properties of Wigner-3$j$ symbols.  Another property involves the correlation structure of the coupling coefficients. To illustrate it, we consider a baseline window function consisting of a rectangular cut of the sky in equatorial coordinates with right ascension  $\pm30\,\text{deg}$, and declination  $\pm15\,\text{deg}$. We apodize the survey window function with an apodization length of $1\,\text{deg}$.
We also  include a point-source mask, consisting of 360 holes of $5\,\text{arcmin}$
radius and an apodization length of $18\,\text{arcmin}$. We define a set of normalized coupling coefficients $\xi^{\alpha \beta}_{\ell_{1} \ell_{2}} = \Xi^{\alpha \beta}_{\ell_{1}  \ell_{2}} ( \Xi^{\alpha \beta}_{\ell_{1}  \ell_{1}} \Xi^{\alpha \beta}_{\ell_{2}  \ell_{2}})^{-1/2} $, analogous to the construction of correlation matrices, and plot them in Fig.~\ref{fig:correlation}.

As can be seen from Fig.~\ref{fig:correlation}, the normalized coupling coefficients exhibit a Toeplitz structure, with each descending diagonal from left to right being close to a constant. Since a $N\times N$, symmetric Toeplitz matrix has elements $A_{ij} = a_{|i-j|}$, for some $N$ coefficients $a_{0},\ldots,a_{N-1}$, we can illustrate the Toeplitz structure by plotting rows of the coupling coefficients, $\xi^{\alpha \beta}_{\ell_{1},\ell_1 + \Delta \ell}$, against $\Delta \ell$. We see from the figure that
for large enough multipoles they are 
nearly indistinguishable. We note that $ \xi^{--}_{\ell_{1}  \ell_{2}}$ departs from Toeplitz the most. However this term, which represents the leakage between E and B modes, is subdominant. The contribution to pseudo-EE from BB is always small and the contribution from EE to pseudo-BB is generally small at high $\ell$.

The fact that the matrices follow this symmetry is extremely encouraging for their efficient computation. If the normalized coupling coefficients were exactly Toeplitz, the computation of all their elements would be reduced to the computation of a single row (or column) of the matrix, turning the computation of the coupling kernels into an $\mathcal{O}(\ell^{2}_{\rm max})$ operation.

However, as can be seen from the right panels of Fig. \ref{fig:correlation}., the Toeplitz symmetry is broken at low and intermediate multipoles. In order to quantify the validity of the approximation, we compute the fractional difference on the inferred pseudo-power spectrum between the exact computation and the Toeplitz approximation:
\ba
\frac{(\Delta \tilde{C}^{TT}_{\ell_{1}} )_{\ell_2} }{\tilde{ C}^{TT}_{\ell_{1}} } =\frac{ |\Xi^{00}_{\rm exact} - \Xi^{00}_{\rm toeplitz} |_{\ell_{1}\ell_{2}}  (2\ell_{2} + 1) C^{TT}_{\ell_2} }{\tilde{C}^{TT}_{\ell_1}} .
\ea
Here, $\Xi^{00}_{\rm exact}$ is the exact coupling kernel, and $\Xi^{00}_{\rm toeplitz}$ is an approximation to $\Xi^{00}_{\rm exact}$ where we assume that the normalized coupling coefficient is exactly Toeplitz,  and where we use the computation of its last column to fill in the rest of the matrix. We  build similar metrics for the polarisation power spectra and show results for the TT and EE power spectra in Fig. \ref{fig:metric}.
The plot displays the magnitudes of the contributions to the fractional errors on the inferred pseudo-power spectra from each matrix element.   Not only does it take into account  the fact that the normalized coupling coefficient are not exactly Toeplitz, but it also weights the difference using the shape of the underlying power spectra. The CMB power spectra generally being very red (in particular in temperature), an error on the coupling element  $\Xi^{00}_{\ell_{1}\ell_{2}} $ for small $\ell_{2}$ may have an important effect on the final spectra.

\begin{figure}
\includegraphics[width=0.85\columnwidth]{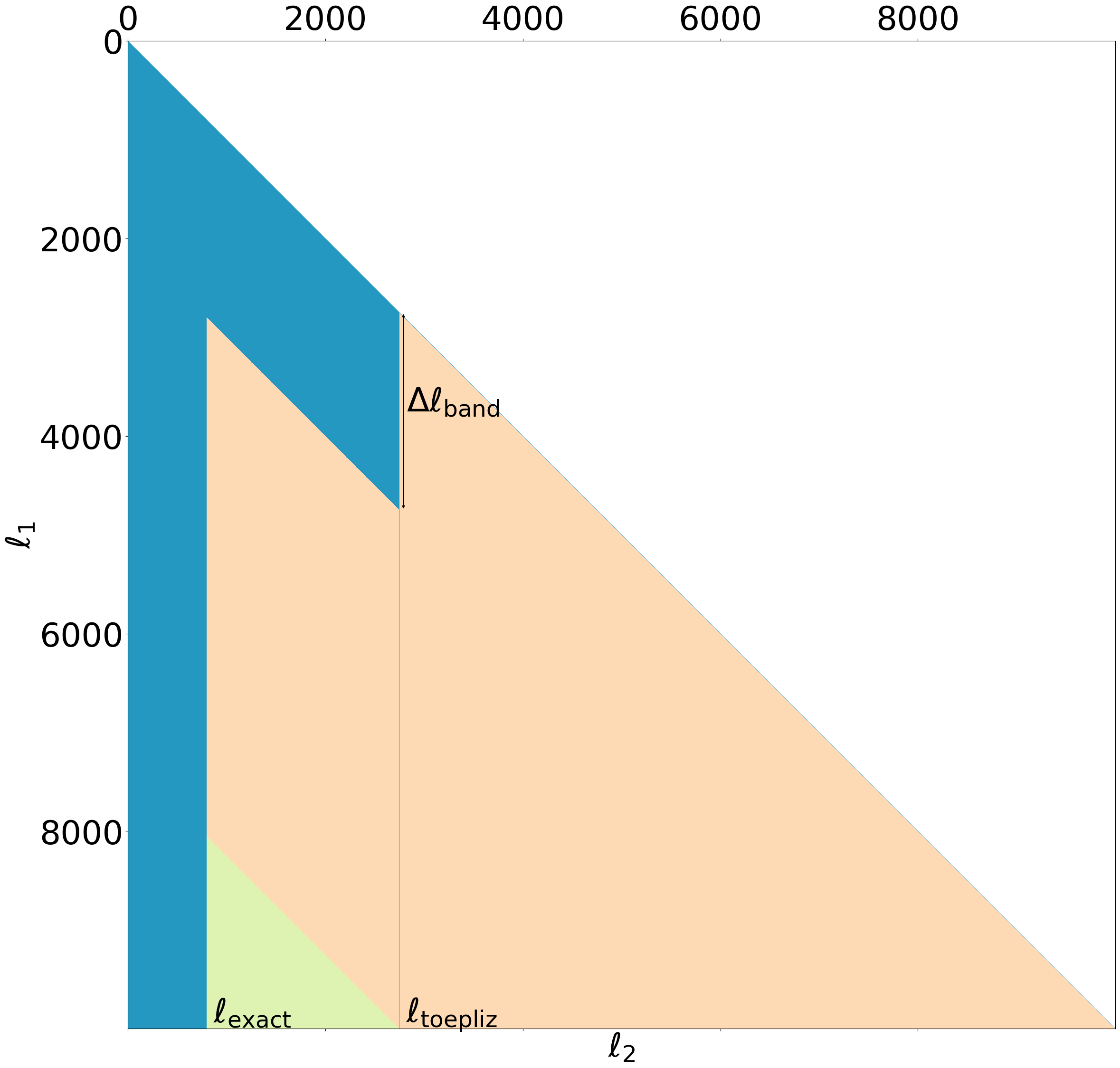}
\caption{Suggested scheme for accurate approximate calculation of the coupling matrices. In blue, we display the elements requiring exact computation. They consist of the diagonal of the matrix, all elements with  $\ell_{2} < \ell_{\rm exact}$ where leakage is important due to the fact that the spectra may be large on large scales, and a band of size $\Delta \ell_{\rm band}$ along the diagonal at low and intermediate multipoles. In orange we show the elements of the matrix that are filled  with the Toeplitz approximation using the single column measured at $\ell_{2} = \ell_{\rm toeplitz}$.  We fill the remaining elements (in green) using the correlation structure measured at $\ell_{\rm exact}$. }
 \label{fig:elements}
\end{figure}

\begin{figure*}
\includegraphics[width=1\textwidth]{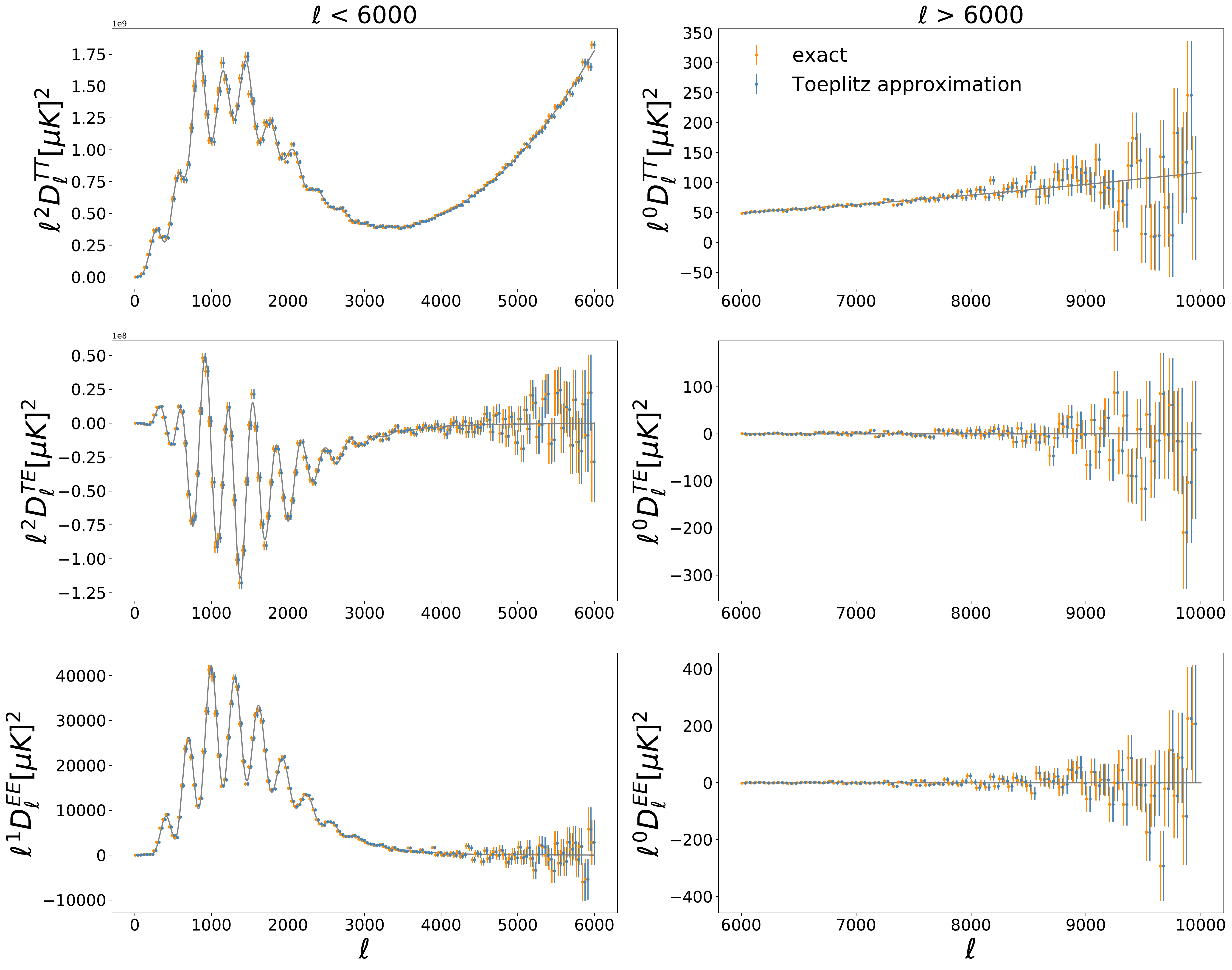}
\caption{Comparison of the estimated TT, TE and EE power spectra using the exact MASTER algorithm (orange) and the Toeplitz approximation (blue) for the baseline survey. We estimate spectra in bands of constant width $\Delta \ell = 40$, and plot the two sets of estimates offset horizontally for clarity.
We adopt the following set of parameters: $\ell_{\rm exact}=800$,   $\Delta\ell_{\rm band}=2000$ and $\ell_{\rm toeplitz}=2750$. The Toeplitz approximation allows for an order-of-magnitude faster computation but yields very precise results. Note that we have split the spectra between $\ell <6000$ and $\ell>6000$, in order to  use different $\ell$ scaling for visualization purposes, and have defined 
$D_\ell \equiv \ell (\ell+1) C_\ell / 2\pi$. The error bars are derived analytically, as described in Sec.~\ref{sec:approx}.}
 \label{fig:spectra}
\end{figure*}

Using this error metric, we identify matrix elements for which the Toeplitz approximation will fail. We find that the elements are located either at low $\ell_{2}$ or in a band along the diagonal at low and intermediate multipoles. This is expected and can be explained both by the redness of the power spectra and by the fact that the coupling coefficients decrease rapidly away from the diagonal. This allows us to define a better approximation scheme for the coupling coefficients than the simple Toeplitz approximation: elements with large error should be computed exactly while the rest of the matrix is filled assuming the Toeplitz approximation.

Our suggestion for the elements of the coupling matrices requiring exact computation is illustrated in Fig.~\ref{fig:elements}. For the rest of the matrix that is filled with the Toeplitz approximation, we suggest using the single column computed exactly at $\ell_{2} = \ell_{\rm toeplitz}$. Since the maximum correlation length that we have measured at $\ell_{\rm toeplitz}$ is $\ell_{\rm max} - \ell_{\rm toeplitz}$ some elements cannot be filled as illustrated by the green triangle in the figure. We fill these remaining elements using the correlation structure measured at $\ell_2 = \ell_{\rm exact}$.

\subsection{Origin of the Toeplitz structure}\label{subsec:origin}

In this subsection, we provide some justification for the approximate Toeplitz structure of the coupling matrices by considering the asymptotic limit of the $3j$-symbols in the limit $\ell_1,\ell_2 \gg \ell_3$. This limit is appropriate if the survey window function is geometrically not too small, the density of any point-source cuts in the survey mask is not too high, and suitable apodization is applied. If these conditions hold, the power spectrum of the survey window function, $\mathcal{W}_{\ell_3}$ (we drop the dependence on frequency and temperature/polarisation mask for simplicity) is very red and, within the vicinity of the diagonal, $\Xi^{\alpha \beta}_{\ell_{1}  \ell_{2}}$ will be dominated by modes with $\ell_3 \ll \ell_1,\ell_2$. 

In the limit $\ell_1,\ell_2 \gg \ell_3$, the $3j$-symbols
may be approximated by (e.g.,~\cite{Varshalovich:1988ye})
\begin{equation}
\left(\begin{array}{ccc}
\ell_1 & \ell_2 & \ell_3 \\
m_1 & m_2 & m_3 
\end{array} \right) \sim \frac{(-1)^{\ell_2+m_2}}{\sqrt{2\ell_2+1}}
d^{\ell_3}_{m_3,\ell_2-\ell_1}(\theta) , 
\end{equation}
where $\cos\theta = - m_2 /\sqrt{\ell_2(\ell_2+1)}$, $d^\ell_{m,m'}(\theta)$ are the reduced Wigner rotation matrices, and we have assumed integer-valued $\ell_i$. If we consider first the temperature case, we have
\begin{equation}
\left(\begin{array}{ccc}
\ell_1 & \ell_2 & \ell_3 \\
0 & 0 & 0 
\end{array} \right)^2 \sim \frac{1}{(2\ell_2+1)}
\left[d^{\ell_3}_{0,|\ell_2-\ell_1|}(\pi/2)\right]^2 ,
\label{eq:asympTT}
\end{equation}
where we have used $d^\ell_{m,m'}(\beta) = (-1)^{\ell-m} d^\ell_{m,-m'}(\pi-\beta)$ to replace $\ell_2-\ell_1$ with its magnitude $|\ell_2-\ell_1|$.
Note that the right-hand side of Eq.~\eqref{eq:asympTT} is not symmetric in $\ell_1$ and $\ell_2$, because of the pre-factor $1/(2\ell_2+1)$, while the left is. However, the fractional asymmetry is at most of $\mathcal{O}(\ell_3/\ell_1)$, so we recover symmetry asymptotically. 
If we form the normalized coupling coefficients, we have
\begin{equation} 
\xi^{00}_{\ell_1 \ell_2} \propto \sqrt{\frac{2\ell_1+1}{2\ell_2+1}} \sum_{\ell_3}
\frac{(2\ell_3+1)}{4\pi} \mathcal{W}_{\ell_3} \left[d^{\ell_3}_{0,|\ell_2-\ell_1|}(\pi/2)\right]^2
\end{equation}
in the limit that all relevant $\ell_3 \ll \ell_1,\ell_2$. Approximating the pre-factor by unity in the vicinity of the diagonal, we recover the Toeplitz structure.

The reasoning is similar for polarisation. In this case, we have
\begin{equation}
\left(\begin{array}{ccc}
\ell_1 & \ell_2 & \ell_3 \\
2 & -2 & 0 
\end{array} \right) \sim \frac{(-1)^{\ell_2}}{\sqrt{2\ell_2+1}}
d^{\ell_3}_{0,\ell_2-\ell_1}(\theta) , 
\end{equation}
where $\cos\theta = 2/\sqrt{\ell_2(\ell_2+1)}$. It follows that $\theta$ is close to $\pi/2$, i.e., $\theta = \pi/2 + \epsilon$, where $\epsilon \approx - 2/\ell_2$. We can now write
\begin{multline}
\frac{\left(1\pm (-1)^{\ell_1+\ell_2+\ell_3}\right)}{2}\left(\begin{array}{ccc}
\ell_1 & \ell_2 & \ell_3 \\
2 & -2 & 0 
\end{array} \right)^2 \sim d^{\ell_3}_{0,\ell_2-\ell_1}(\theta) \\
\times \frac{1}{(2\ell_2+1)} \frac{1}{2}
\left[d^{\ell_3}_{0,\ell_2-\ell_1}(\theta) \pm d^{\ell_3}_{0,\ell_2-\ell_1}(\pi-\theta)\right] ,
\label{eq:asympevenodd}
\end{multline}
where we have used
\begin{align}
(-1)^{\ell_1+\ell_2+\ell_3} \left(\begin{array}{ccc}
\ell_1 & \ell_2 & \ell_3 \\
2 & -2 & 0 
\end{array} \right) &= \left(\begin{array}{ccc}
\ell_1 & \ell_2 & \ell_3 \\
-2 & 2 & 0 
\end{array} \right) \nonumber \\
&\sim \frac{(-1)^{\ell_2}}{\sqrt{2\ell_2+1}}
d^{\ell_3}_{0,\ell_2-\ell_1}(\pi - \theta) ,
\end{align}
which follows from the symmetry properties of the $3j$-symbols.
Taylor expanding the right-hand side of Eq,~\eqref{eq:asympevenodd}, at leading order in $\epsilon$ we have
\begin{equation}
\left(\begin{array}{ccc}
\ell_1 & \ell_2 & \ell_3 \\
2 & -2 & 0 
\end{array} \right)^2 \sim \frac{1}{(2\ell_2+1)}
\left[d^{\ell_3}_{0,|\ell_2-\ell_1|}(\pi/2)\right]^2 
\end{equation}
for $\ell_1+\ell_2+\ell_3$ even, and
\begin{equation}
\left(\begin{array}{ccc}
\ell_1 & \ell_2 & \ell_3 \\
2 & -2 & 0 
\end{array} \right)^2 \sim \frac{\epsilon^2}{(2\ell_2+1)}
\left[\left.\frac{ {\mathrm d}\,d^{\ell_3}_{0,|\ell_2-\ell_1|}(\beta)}{{\mathrm d} \beta}\right|_{\pi/2}\right]^2 
\label{eq:minus3jsq}
\end{equation}
for $\ell_1+\ell_2+\ell_3$ odd. Apart from the pre-factors, the right-hand side of these expressions are Toeplitz. Note from Eq.~\eqref{eq:minus3jsq} that the contribution to $\Xi^{--}_{\ell_{1}  \ell_{2}}$ at $\ell_3$ is suppressed by $\mathcal{O}(\ell_3^2/\ell_2^2)$ compared to $\Xi^{++}_{\ell_{1}  \ell_{2}}$. This behaviour arises since the mixing of E- and B-modes due to the mask depends on derivatives of the mask. Indeed, for smooth masks the normalization of the mixing kernel, $\sum_{\ell_2} \Xi^{--}_{\ell_1 \ell_2}$, is determined by the average squared gradient of the mask~\cite{2005MNRAS.360..509C}.

\section{Tests of the Toeplitz approximation}\label{sec:results}

\subsection{Baseline survey}

For our first set of tests we use the baseline window function defined in Sec.~\ref{sec:approx}. This window function covers $1800\,\text{deg}^2$
of the sky. Recall that the point-source mask consists of 360 holes with $5\,\text{arcmin}$ radius and an apodization length of $18\,\text{arcmin}$, resulting in a source mask density of $0.2\,\text{deg}^{-2}$. 
We use the same survey window function for temperature and polarisation. We adopt $0.5\,\text{arcmin}$ pixels in both right ascension and declination.

\begin{figure}
\includegraphics[width=\columnwidth]{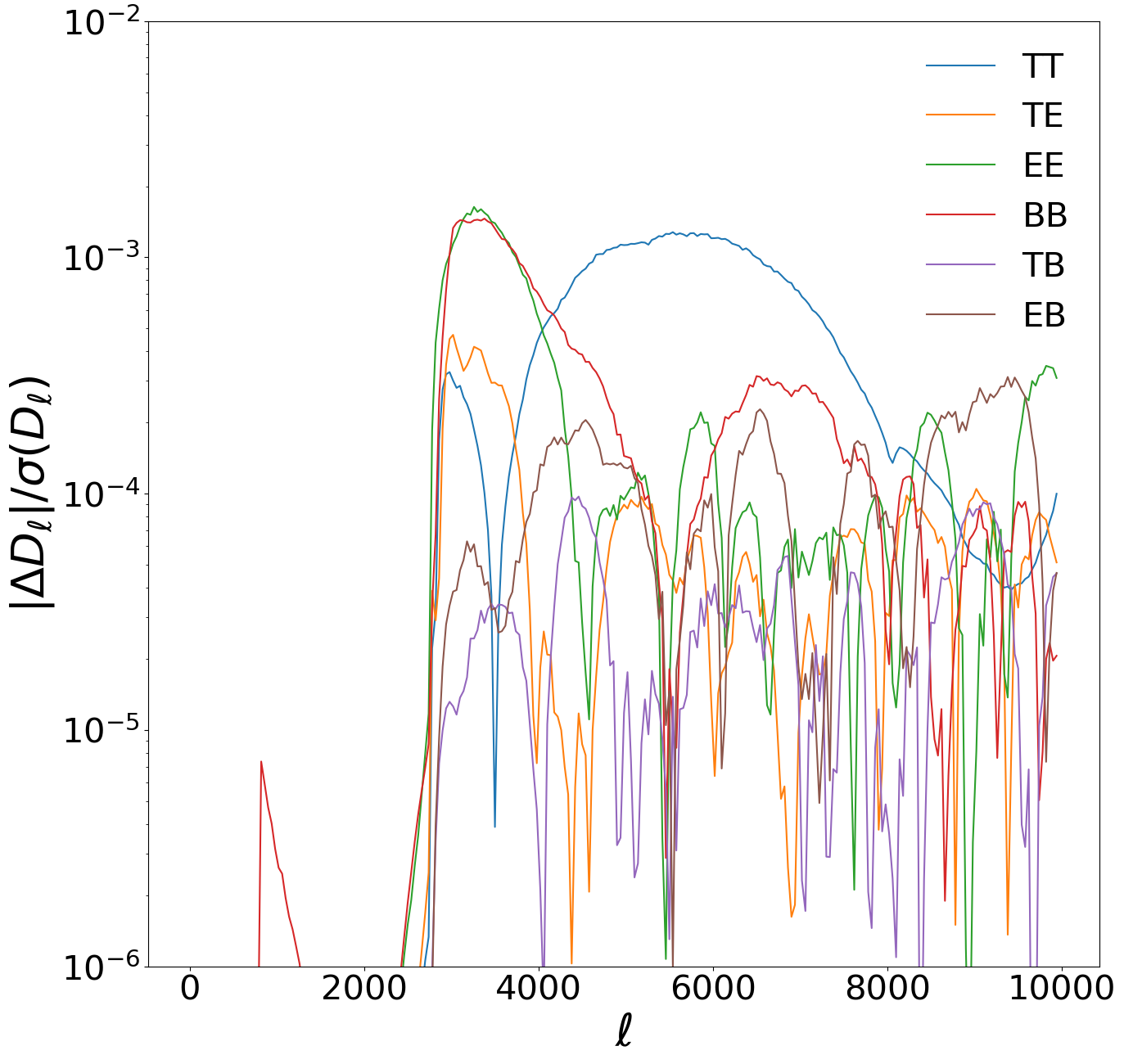}
\caption{Difference between the power spectra estimated using the exact MASTER algorithm and the Toeplitz approximation in units of the standard deviations of bandpowers of constant width $\Delta \ell =40$.
The baseline survey is adopted, as in Fig.~\ref{fig:spectra}, and the same set of approximation parameters.
We see that the effect of the approximation is at most a fraction of a percent of the error bars.}
 \label{fig:residual}
\end{figure}

\begin{figure}
\includegraphics[width=\columnwidth]{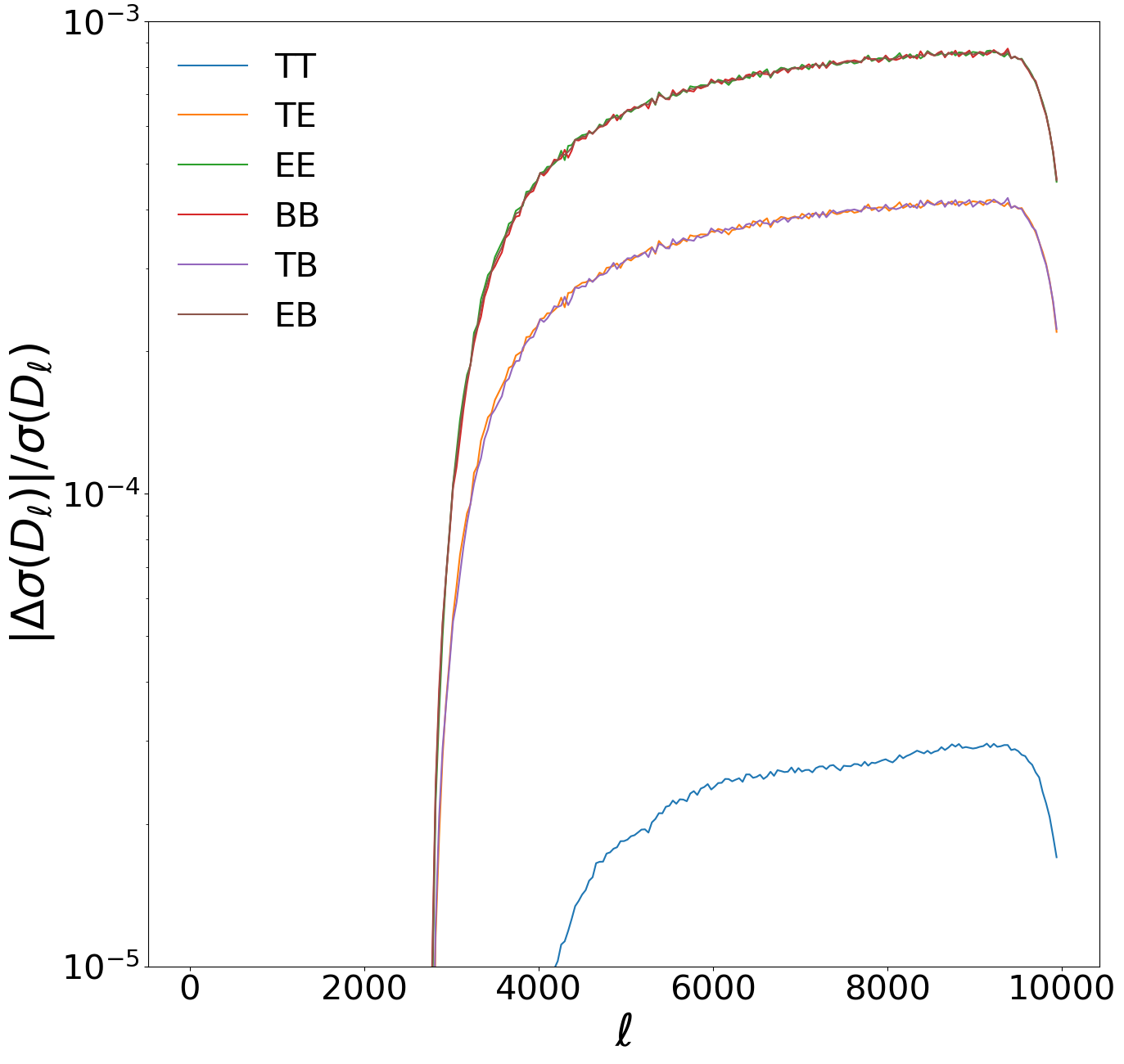}
\caption{Fractional differences on the estimated power spectrum errors using the exact MASTER algorithm and the Toeplitz approximation. The effect of the approximation on the computation of the power spectrum errors is at most of order $0.1\,\%$. The (baseline) survey and approximation parameters follow Fig.~\ref{fig:spectra}.
The spectra are arranged in three groups:  polarisation only;  cross temperature-polarisation; and temperature only. The Toeplitz approximation performs best for the errors on the temperature power spectrum, as expected from Fig.~\ref{fig:metric}.  We use bins of constant width $\Delta \ell =40$. }
 \label{fig:errors}
\end{figure}

We generate CMB simulations from a set of temperature and polarisation power spectra compatible with the {\it Planck} legacy cosmology \cite{2018arXiv180706205P}. In temperature, we also simulate extragalactic foreground emission with amplitudes compatible with  measurements from the Atacama Cosmology Telescope \cite{2013JCAP...07..025D} at $93\,\text{GHz}$. We convolve the simulations with a $2.3\,\text{arcmin}$ Gaussian beam corresponding to the angular resolution of the 93\,GHz channel of a 6\,m telescope.
We generate two separate observations of each sky simulation by drawing two independent white-noise realizations with standard deviation $\sigma_{\rm T} = 2\sqrt{2}\,\mu\text{K\,arcmin}$ in temperature and $\sigma_{\rm pol} = \sqrt{2} \sigma_{\rm T}$ in polarisation, for a coadded noise of $2 \mu\text{K\,arcmin}$ in temperature.
These simulations are simplistic but are representative of the high-$\ell$  goal sensitivity of next-generation surveys \cite{2019JCAP...02..056A, 2016arXiv161002743A}.

We compute the cross-power spectra between the two simulated realizations both using the exact MASTER algorithm and the Toeplitz approximation with parameters $\ell_{\rm exact}=800$,   $\Delta \ell_{\rm band}=2000$ and $\ell_{\rm toeplitz}=2750$. We bin the measured spectra and coupling matrices into bins of constant width $\Delta \ell = 40$ before inverting to obtain binned estimates of the true spectra. Our results are shown in Fig. \ref{fig:spectra} for the TT, TE and EE spectra. As can be seen, the agreement between the exact and approximated computation is excellent.
We also show the amplitude of the differences between the exactly- and approximately-computed estimates of all the spectra, as a fraction of the statistical errors in the binned spectra, in Fig.~\ref{fig:residual}. We find that in this configuration, the difference between the two power spectra never exceed $1\,\%$ of the statistical error. {Note that we include the $TB$ and $EB$ spectra here, which are expected to vanish if the distribution of primordial fluctuations is invariant under parity.}
Finally we show the fractional error on our analytic computation of the error bars on the power spectra in Fig.~\ref{fig:errors}. For the survey and approximation parameters adopted,  we find that the approximation leads to fractional errors on the power spectrum uncertainties of order $10^{-3}$. We find that the spectra fall into three groups:  polarisation only (EE/EB/BB);  cross temperature-polarisation (TE/TB); and temperature only (TT).
The Toeplitz approximation performs best for the TT spectrum as expected from Fig.~\ref{fig:metric}.

These numbers suggest that the difference between the results of an exact analysis and one made with the Toeplitz approximation would be essentially indistinguishable for this baseline survey. Most of the analysis in this case could be done using the Toeplitz approximation without any relevant loss of accuracy.

\begin{table}[t!]
\caption{\small Summary of the benchmark for the baseline case. Times are given in second.
}
\centering
\begin{tabular}{lccc}
\hline
\hline
     &  Nersc (Haswell)   &   Niagara & MacBook Pro \\
\hline
Toeplitz   & 32.80 $\pm$ 0.05  s  & 18.55 $\pm$ 0.13 s    & 71.36  $\pm$ 1.24 s \\
Exact     & 332.02  $\pm$ 0.04 s  &  203.63 $\pm$ 0.45 s  & 885.09 $\pm$ 18.06 s  \\
\hline
\end{tabular}
\label{tab:data}
\end{table}

\begin{figure*}
\includegraphics[width=0.95\columnwidth]{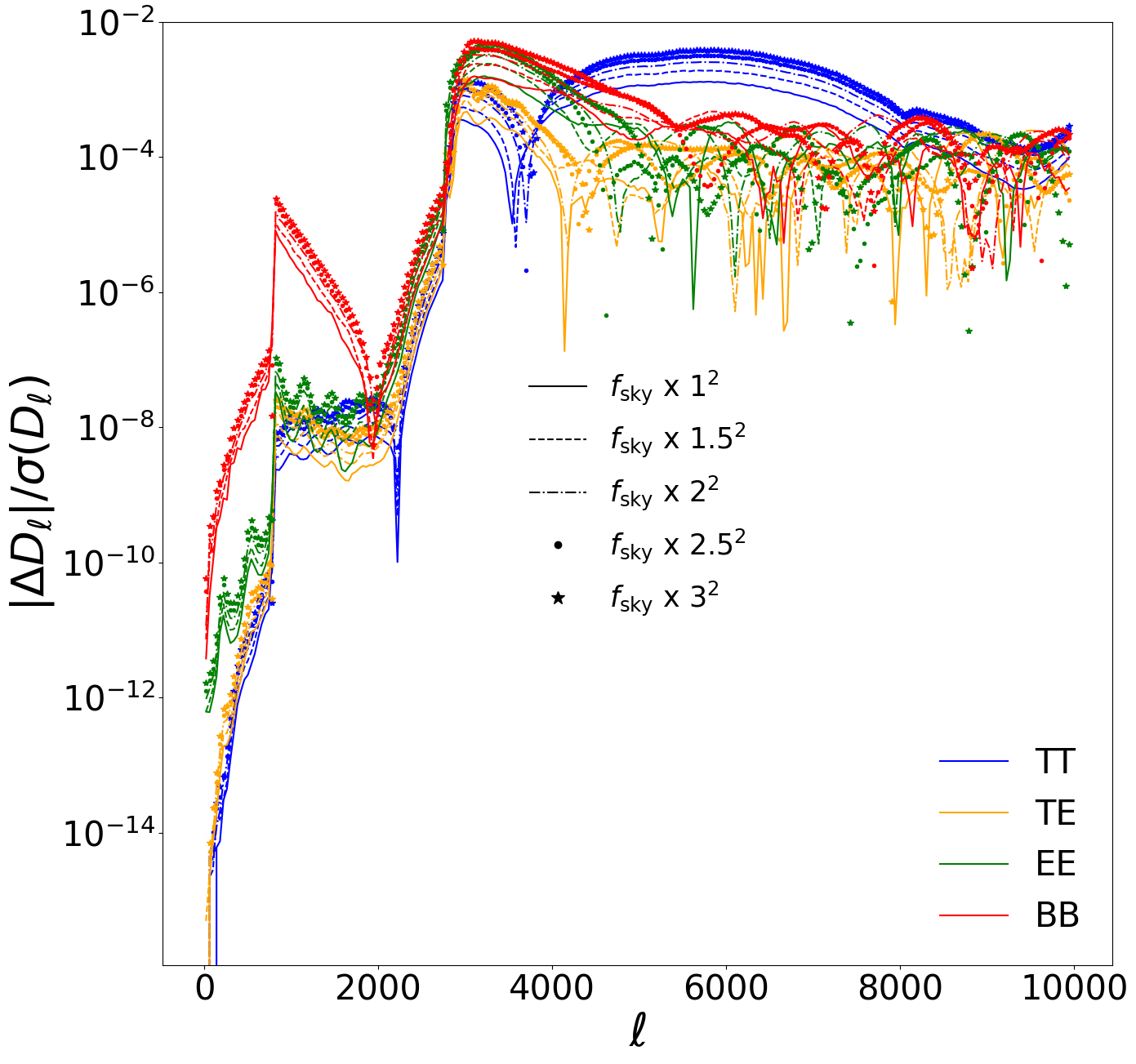}
\includegraphics[width=0.95\columnwidth]{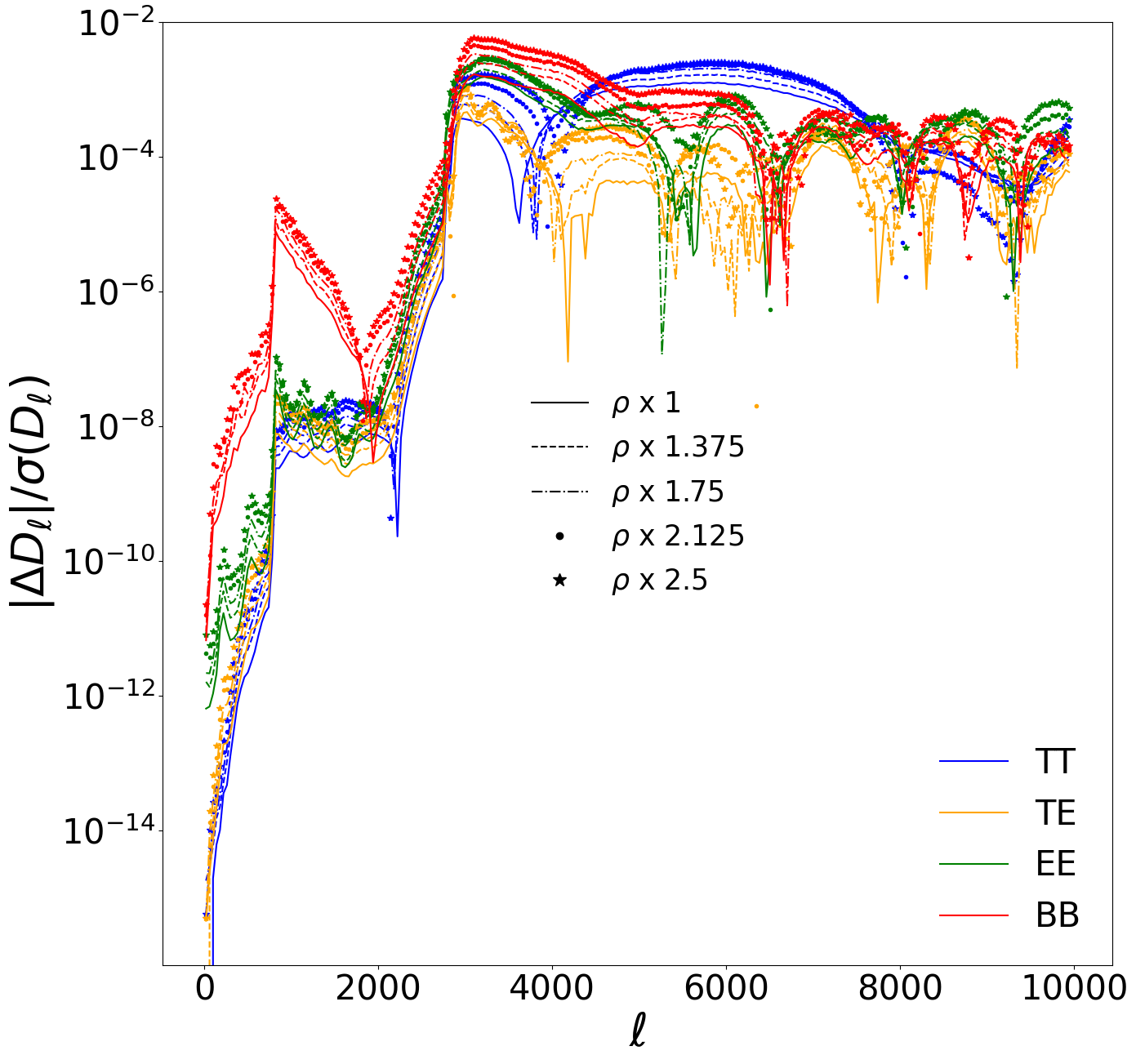}
\includegraphics[width=0.95\columnwidth]{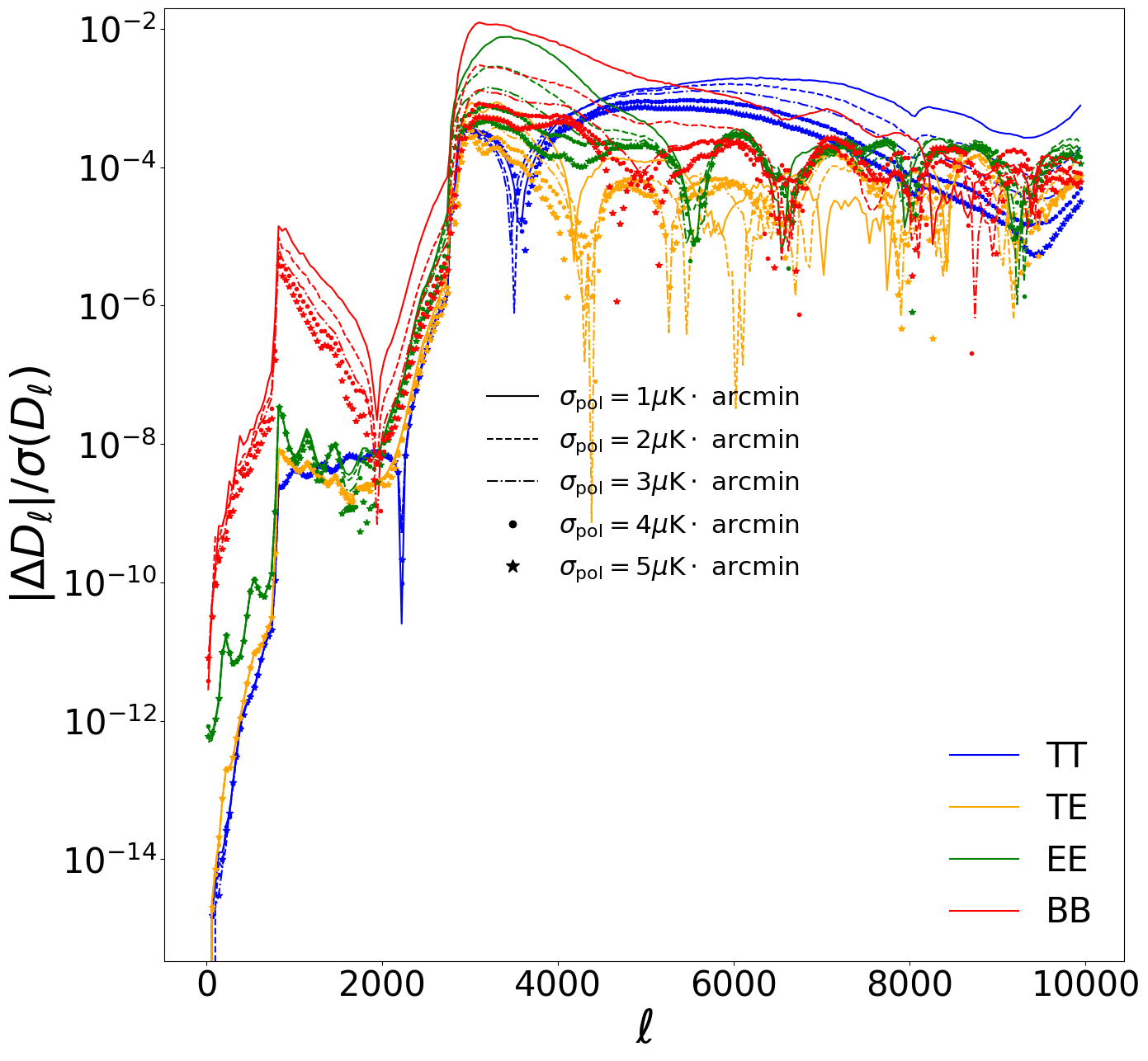}
\includegraphics[width=0.95\columnwidth]{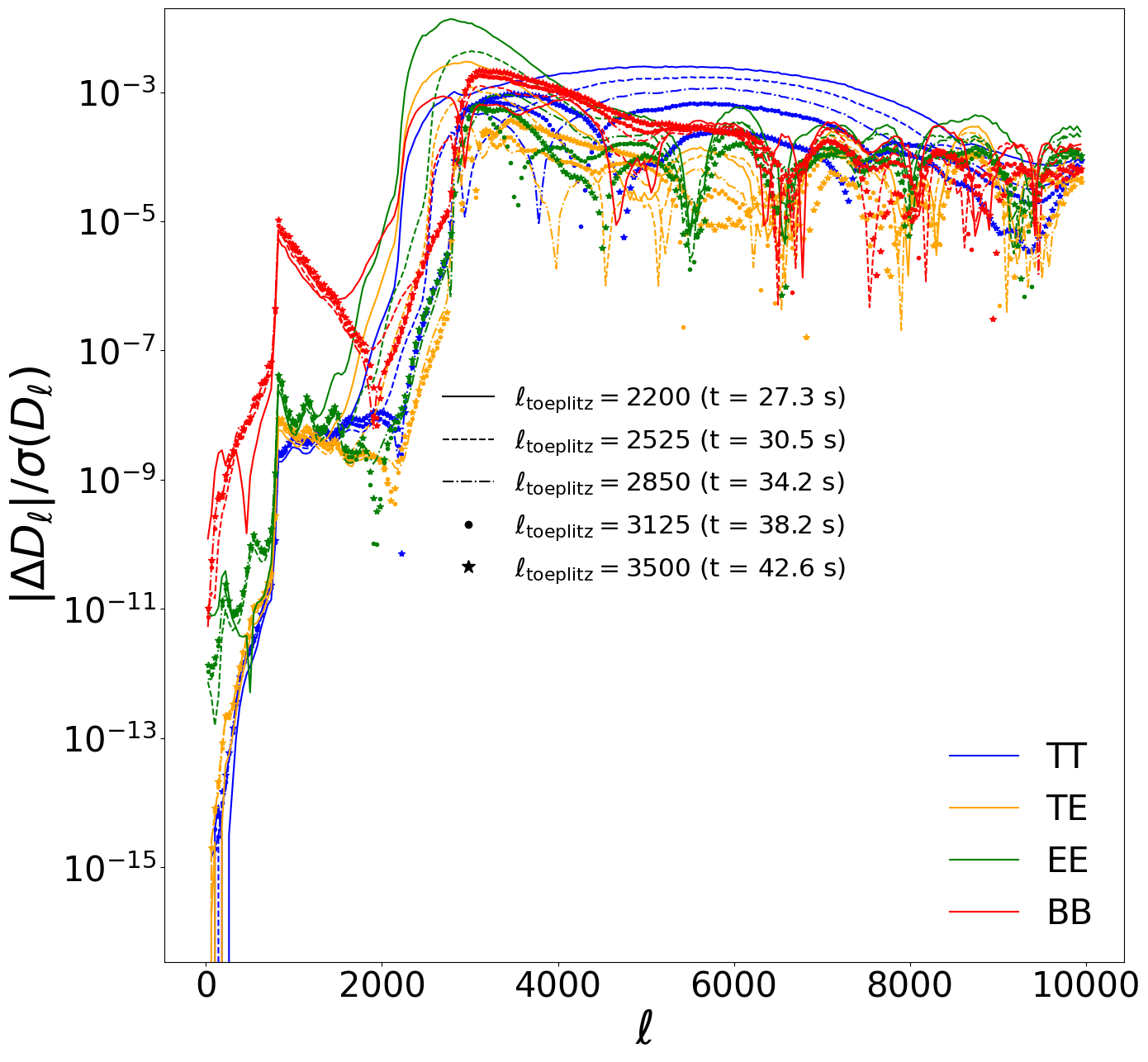}
\includegraphics[width=0.95\columnwidth]{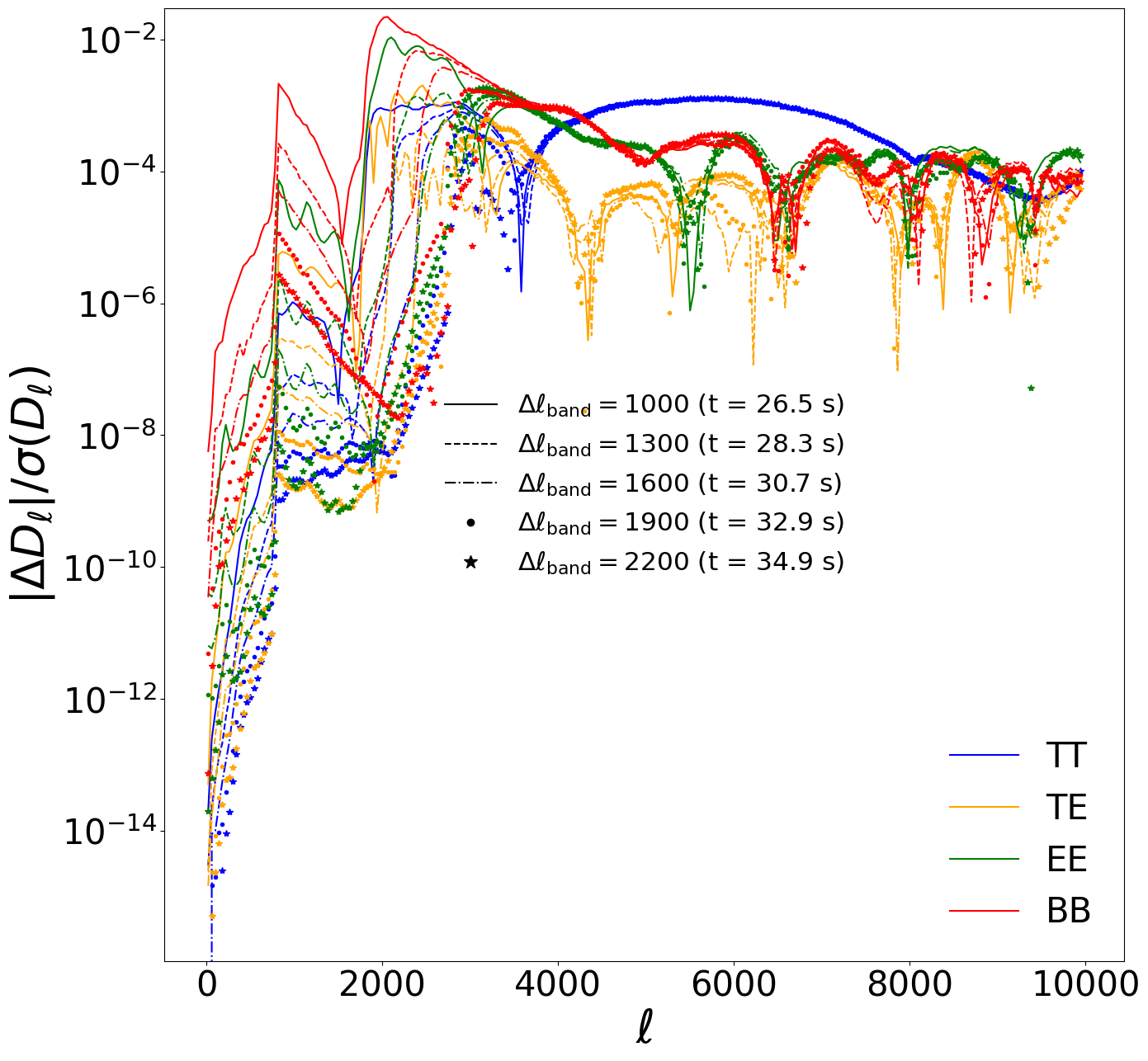}
\includegraphics[width=0.95\columnwidth]{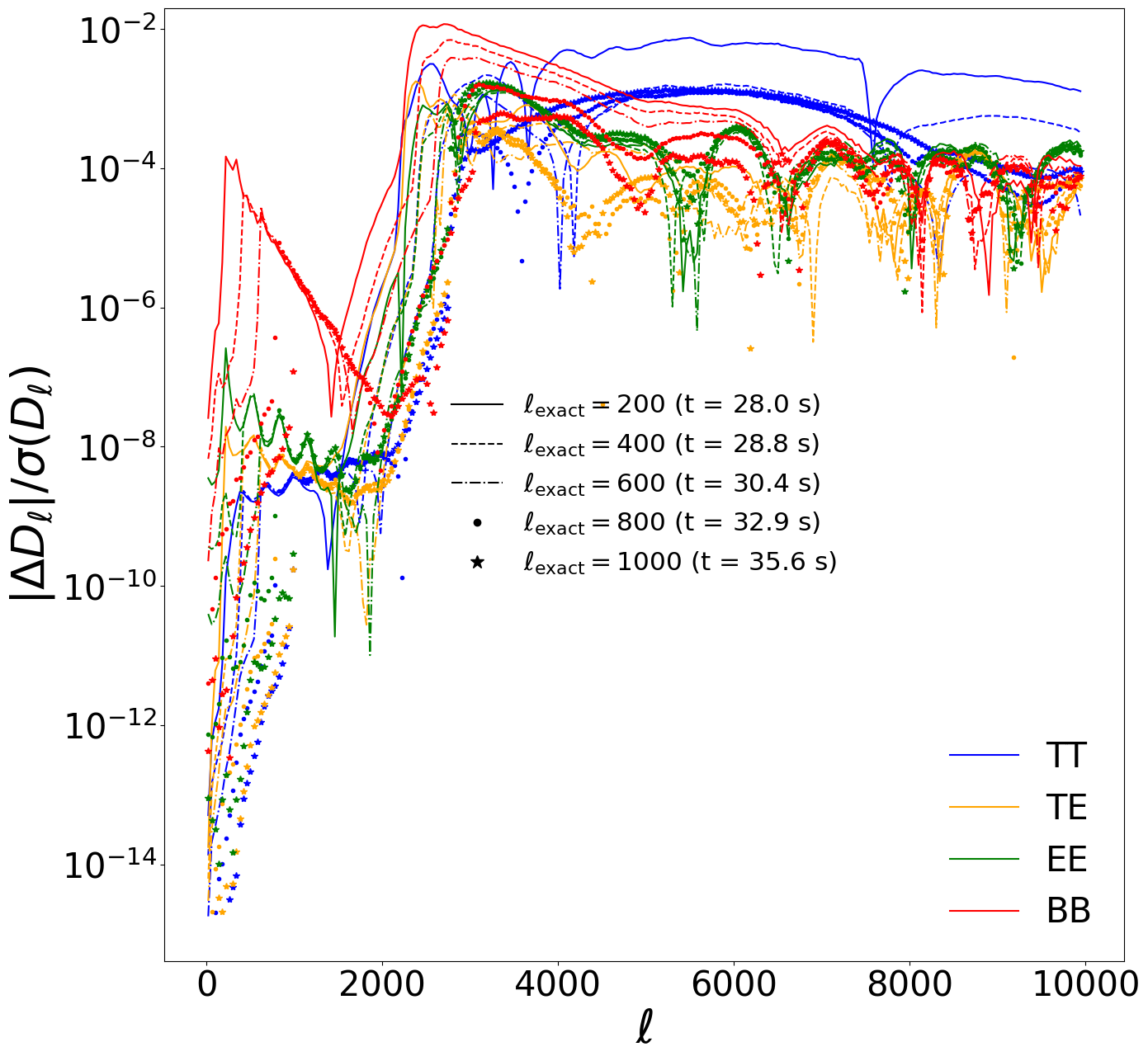}
\caption{ Accuracy of the Toeplitz approximation as a fraction of the bandpower errors, as in Fig.~\ref{fig:residual}, for different sky fractions (top left, as a multiple of the baseline survey), point-source mask density (top right, also as a multiple of the baseline survey density),  noise levels (middle left) and the different multipoles parametrizing the Toeplitz approximation. For clarity we focus on the TT, TE, EE and BB spectra. The different behaviors are described in the text. The computation times quoted are measured on a single NERSC (Haswell) compute node. In the middle left panel, we use $\sigma_{\rm T}= \sqrt{2} \sigma_{\rm pol}$.
}
 \label{fig:tests}
\end{figure*}

We benchmark the code on three different systems. The first one is a NERSC Haswell compute node with two sockets, each populated with a $2.3\,\text{GHz}$, 16-core 
Intel Xeon Processor E5-2698 v3. The second one is a Niagara node with a $2.4\,\text{GHz}$, 40-core Intel Xeon  Gold 6148 CPU processor. Finally, we use a MacBook Pro laptop with 8 cores at $2.4\,\text{GHz}$. The results are presented in Table~\ref{tab:data}.  Uncertainties in running time are computed from 10 different runs of the code. We find speed ups in the computation of the coupling coefficients by factors in the range 10--12.

This reduction in compute time is expected. For the baseline survey, we are computing $23\,\%$ of the coupling matrix elements exactly. However the computation time of each element differs since the Wigner-$3j$ symbols are non-zero only for $ |\ell_{1} - \ell_{2}| \leq \ell_{3} \leq  \ell_{1}+ \ell_{2} $. Not having to compute elements at high $\ell_{1}$ and $\ell_{2}$ explains how we can compute $23\,\%$ of the matrix elements exactly in only 8--10\,\% of the time.

For completeness, we note that we have assumed a form for the analytical errors of the BB spectrum that neglects contributions from E-to-B leakage. The E-to-B leakage
is important on large angular scales and can be reduced, for example, by using the pure-B-mode formalism \cite{ 2007PhRvD..76d3001S}. On these scales, however, the Toeplitz approximation becomes irrelevant since the exact computation is easily tractable. The problem of optimal estimation of the B-mode power spectrum and the analytical computation of its associated errors is disconnected from the problem of speeding up the computation of power spectra of high-angular-resolution CMB maps.

\subsection{Beyond the baseline}

\subsubsection{Effect of sky fraction}

The baseline survey has a sky fraction of around $f_{\rm sky} =4\,\%$. In this subsection we investigate the effect of increasing it keeping constant the density of point-source holes in the mask ($0.2\,\text{deg}^{-2}$).
The results are displayed in the top-left panel of Fig \ref{fig:tests}. We find that the errors made in the approximation are nearly independent of $f_{\rm sky}$, and, since the errors on the power spectra scale as $ f_{\rm sky}^{-1/2}$, we find that the maxima of  $\Delta D_{\ell}/\sigma(D_\ell)$ scale roughly as  $f_{\rm sky}^{1/2}$. For sky fractions around  $36\,\%$, which is around the maximum sky fraction covered by next-generation ground-based surveys, $\Delta D_{\ell}/\sigma(D_\ell)$ stays well below $1\,\%$ for each bin of the power spectra (for $\Delta \ell = 40$).

\subsubsection{Effect of the point-source mask density}

The point-source mask density of the baseline window function is $\rho = 0.2\,\text{deg}^{-2}$, corresponding to 360 masked sources for our $1800\,\text{deg}^2$ survey. The source mask density can be thought as a metric for the mask complexity, with the high-$\ell$ part of $\mathcal{W}_\ell$ increasing linearly with $\rho$.
Here, we investigate the accuracy of the Toeplitz approximation for different source mask densities, ranging from our baseline case up to $\rho = 0.5\,\text{deg}^{-2}$.
Changing the source mask density changes $f_{\rm sky}$, but only by $5\,\%$ in the range $\rho = (0.2$--$0.5)\,\text{deg}^{-2}$, ensuring that $\sigma(D_\ell)$ is nearly constant in this test. The variation seen in the top-right panel of Fig.~\ref{fig:tests}  is therefore due to the approximation becoming less accurate with a more complex mask, as expected given the increase in small-scale power in $\mathcal{W}_\ell$.
We find that even for a source mask density $\rho = 0.5\,\text{deg}^{-2}$, the Topelitz approximation leads to systematic errors in the power spectra below 1\,\% of $\sigma(D_\ell)$ for each bin ($\Delta \ell = 40$).

\subsubsection{Noise level}

 We investigate the effect of changing the white-noise level of the survey, ranging from a case with noise $\sigma_{\rm pol}= \sqrt{2} \sigma_{\rm T}= 5\,\mu\text{K\,arcmin}$ in polarisation to a case with $\sigma_{\rm pol}=1\,\mu\text{K\,arcmin}$, the latter corresponding to the target depth for CMB-S4~\cite{2016arXiv161002743A}. 
The goal of this test is to show that even for futuristic noise levels, the effect of 
errors in the Toeplitz approximation is still small compared to the bandpower random errors; see the middle-left panel of Fig.~\ref{fig:tests}.
We also note that, if needed, $\ell_{\rm toeplitz}$ can be adjusted to accomodate the precision requirements of each individual survey.

\subsubsection{Effect of $\ell_{\rm toeplitz}$, $\ell_{\rm exact}$ and $\ell_{\rm band}$}

The definitions of the parameters $\ell_{\rm toeplitz}$, $\ell_{\rm exact}$ and $\ell_{\rm band}$, which enter the Toeplitz approximation, were given in Fig.~\ref{fig:elements}.
The multipole $\ell_{\rm toeplitz}$ specifies the multipole above which we use the Toeplitz approximation for every element of the matrix but the diagonal.  In the middle-right panel of Fig.~\ref{fig:tests}  we show the effect of varying $\ell_{\rm toeplitz}$ on the precision of the approximation as well as the associated computing time. We find in, particular, that the E-mode power spectrum is very sensitive to the value of $\ell_{\rm toeplitz}$, with an order-of-magnitude difference in accuracy as $\ell_{\rm toeplitz}$ varies in the range 2200--3500. This behavior is consistent with the error structure shown in the right panel of Fig.~\ref{fig:metric}.
We show the effect of $\Delta \ell_{\rm band}$ and $\ell_{\rm exact}$ in the bottom panels of Fig.~\ref{fig:tests}; we find that they affect mainly the accuracy of the BB power spectrum.

\section{Conclusion}\label{sec:conclu}

In this paper we have presented a Toeplitz approximation that allows fast computation of the mode-coupling matrices, a crucial but slow step in the estimation of angular power spectra and covariance matrices from high-resolution CMB maps. We show that it leads to more than an order-of-magnitude reduction in computation time when analyzing data to a maximum multipole $\ell_{\text{max}} = 10^4$, with very little loss of accuracy. We believe our results will be useful for current and next-generation ground-based CMB experiments. 
  
\section*{Acknowledgments}
TL thanks Steve Choi and David Alonso for interesting discussions. This research used resources of the National Energy Research Scientific Computing Center (NERSC), a U.S. Department of Energy Office of Science User Facility operated under Contract No. DE-AC02-05CH11231.  Some computations were performed on the Niagara \cite{2019arXiv190713600P} supercomputer at the SciNet HPC Consortium. SciNet is funded by: the Canada Foundation for Innovation; the Government of Ontario; Ontario Research Fund - Research Excellence; and the University of Toronto. Flatiron Institute is supported by the Simons Foundation. AC acknowledges support from the STFC (grant numbers ST/N000927/1 and ST/S000623/1).

 \bibliography{draft}

\end{document}